\documentclass[reprint,amsmath,amssymb,aps,floatfix,prd]{revtex4-2}
\pdfoutput=1
\usepackage{graphicx} 
\usepackage{dcolumn} 
\usepackage{bm} 
\usepackage{siunitx}
\usepackage{subcaption}
\usepackage{caption}
\usepackage{float}
\usepackage{hyperref}
\usepackage{multirow}
\usepackage[dvipsnames]{xcolor}
\hypersetup{colorlinks = false, linkcolor={blue}, citecolor={red}, urlcolor= {blue}}  
\begin{document}
\title{Light front analysis towards the study of thermalisation in pp collisions at $\sqrt{s}= 13$ TeV}
\author{Rahul Ramachandran Nair}
\email{physicsmailofrahulnair@gmail.com}
\thanks{\\Orcid ID: 0000-0001-8326-9846}
\affiliation{National Centre For Nuclear Research, 02-093 Warsaw, Poland}
\altaffiliation{The work was performed while the author was affiliated with the National Centre For Nuclear Research, Warsaw, Poland.\\Present address: Sreeraj Bhavan, Karukachal PO, Kottayam (Dist), Kerala, India- 686540} 
\date{\today}
\begin{abstract}
An analysis involving the light front variables of inclusively produced hadrons in proton-proton (pp) collisions to study the thermalisation and formation of a QGP-like medium is presented in this paper. Two schemes of analysis are discussed and performed with inclusively produced $\pi^{\pm}$, $K^{\pm}$ and $p(\bar{p})$ in the central pp collisions at $\sqrt{s} = 13$ TeV simulated using the PYTHIA 8 event generator. It is shown that a group of $\pi^{\pm}$ and $K^{\pm}$ falling inside a paraboloid defined by a certain constant value of the light front variable in their respective phase space follows the Bose-Einstein statistics while the $p(\bar{p})$ inside the similar paraboloid in its phase space follows the Fermi-Dirac statistics. It is also shown that for the $\pi^{\pm}$, $K^{\pm}$ and $p(\bar{p})$ with transverse momentum $p_{T} < 1$ GeV in these collisions, a polynomial in $p_T$ can be constructed for a specific light front variable of the particle $\zeta_c (p_T)$ such that the $\pi^{\pm}$ and $K^{\pm}$ with their light front variable $\zeta^{\pm} > \zeta_c (p_T)$ follows Bose-Einstein statistics while the $p(\bar{p})$ with  $\zeta^{\pm} > \zeta_c (p_T)$ follows the Fermi-Dirac statistics. The analysis can serve as a baseline for an investigation towards the QGP formation and thermalisation in the pp collisions at LHC.
\end{abstract}
\maketitle
\section{Introduction} 
The investigations regarding the possible formation of quark-gluon plasma (QGP) like medium and thermalisation in proton-proton (pp) collisions are of prime interest. The reliability of treating the pp collisions as a benchmark for the comparison of the results from nucleus-nucleus collisions relies heavily on the outcome of such studies. The observations like strangeness enhancement, which is considered as a signature effect of deconfinement, in high-multiplicity pp collisions as reported in \cite{doi:10.1142/S0217751X17300241} by the ALICE experiment \cite{Collaboration_2008, Adam2017} at Large Hadron Collider (LHC) suggests that a naive benchmarking of pp collisions for interpreting the results of heavy-ion collisions might not be a wise idea. Similarly, the experimental evidence of collectivity in pp collisions similar to the observations in pPb and PbPb collisions at the CMS experiment at LHC \cite{2017193} invites a careful study of the scenario. Given that the hydrodynamical models, where a local thermal equilibrium is assumed, can describe the experimental observables at sufficiently low $p_T$ \cite{PhysRevLett.113.252301, PhysRevLett.111.172303} and the ability of these models to describe the $p_T$ dependence of the transverse elliptical flow (which is interpreted as evidence of QGP formation in heavy-ion collisions \cite{STRICKLAND2015}), the investigation of QGP-like a medium in central pp collisions have significant consequences in the interpretation of the results from heavy-ion collisions \cite{Nachman2018}.  \par In this paper, an approach involving the light front variables \cite{RevModPhys.21.392} is presented towards the study of thermalisation and QGP formation in pp collisions with the PYTHIA 8 model of ultra-relativistic hadron-hadron collisions \cite{Sj_strand_2006, SJOSTRAND2008852}. The organisation of this paper is as follows. We will first discuss the idea of light front analysis to study thermalisation in hadron-hadron collisions. The details of the event generation and tuning of PYTHIA 8,  the trigger criterion are presented in the following section. The two schemes of analysis performed in the current work with their results are presented afterwards. In the final section, we will comprehensively discuss the results and possible implications of our analysis.
\section{The light front analysis}
The light front variable was proposed in \cite{Garsevanishvili78, Garsevanishvili79} for studying the particle production in hadron-hadron and nucleus-nucleus interactions. The proposed scale and Lorentz invariant variable has the following form in the centre of the mass frame
\begin{equation} \label{EqnXi}
\xi^{\pm} = \pm \frac{E + |p_{z}|}{\sqrt{s}}
\end{equation}
where, $s$ is the Mandelstam variable, $p_{z}$ is the z-component of the momentum and E is the energy of the particle. The positive and negative signs in Eq. \eqref{EqnXi} corresponds to the two hemispheres. In order to enlarge the region of smaller $\xi$ values in the $\xi^{\pm}$ distribution for the convenience of analysis, a variable $\zeta^{\pm}$ can be defined as follows:
\begin{equation} \label{EqnZeta}
\zeta^{\pm} = \mp\ln(\xi^{\pm})
\end{equation}
The analysis scheme based on these variables were performed for hadron-hadron and nucleus-nucleus interactions in \cite{Amaglobeli99, Djobava03, Chkhaidze2006}. The value close to the experimental maximum of the $\xi^{\pm}$ distribution denoted by $\tilde{\xi}^{\pm}$ for $\pi^{\pm}$ was found to be of significant interest from those studies. The particles with their $\xi^{\pm} < \tilde{\xi}^{\pm}$ were observed to have a comparatively isotropic polar angle distribution with respect to the corresponding distributions of those particles having their $\xi^{\pm} > \tilde{\xi}^{\pm}$. The slopes of the $p_{T}^{2}$ distribution for these two group of $\pi^{\pm}$ also were found to have different slopes. From these curious observations, it was hypothesised that a thermalisation has been reached in the region of the phase space of $\pi^{\pm}$ with $\xi^{\pm} < \tilde{\xi}^{\pm}$. A constant value of the $\xi^{\pm}$ defines the following paraboloid in the phase space:
\begin{equation}
p_{z} = \frac{p_T^2+m^2- (\tilde{\xi}^{\pm}\sqrt{s})^2}{-2\tilde{\xi}^{\pm}\sqrt{s}}
\label{paraboloid}
\end{equation}
The invariant differential cross-section in terms of the light front variables can be written as follows:
\begin{equation}\label{cross}
E\frac{d \sigma}{d \textbf{p}} = \frac{\xi^{\pm}}{\pi}\frac{d^2\sigma}{d\xi^{\pm}dp_T^2} = \frac{1}{\pi}\frac{d^2\sigma}{d\zeta^{\pm}dp_T^2}
\end{equation}
In order to explore the hypothesis of thermalisation of $\pi^{\pm}$ with $\xi^{\pm} < \tilde{\xi}^{\pm}$ or equivalently $\zeta^{\pm} > \tilde{\zeta}^{\pm}$ a set of equations were constructed in \cite{Amaglobeli99, Djobava03, Chkhaidze2006} using the relation in Eq.\eqref{cross}. As per this scheme, the $\zeta$ distribution of particles with $\zeta^{\pm} > \tilde{\zeta}^{\pm}$ should be describable with the following expression:
\begin{equation}
\frac{dN}{d\zeta^{\pm}} \sim \int_0^{p_T^2(max)} E f(E)dp_T^2
\label{ZetaInt}
\end{equation}
the $cos(\theta)$ distribution of particles with $\zeta^{\pm} > \tilde{\zeta}^{\pm}$ must follow
\begin{equation}
\frac{dN}{dcos(\theta)} \sim \int_0^{p_{max}} f(E) p^2dp
\label{CosInt}
\end{equation}
and the $p_T^2$ distribution of particles with $\zeta^{\pm} > \tilde{\zeta}^{\pm}$ should be describable with
\begin{equation}
\frac{dN}{dp_T^2} \sim \int_0^{p_{z,max}} f(E)dp_{z}
\label{PtSqInt}
\end{equation}
while the $f(E)$ in these expressions takes the Bose-Einstein (for $\pi^{\pm}$ and $K^{\pm}$) or Fermi-Dirac (for $p(\bar{p})$) form given by
\begin{equation}\label{eqboltz}
f(E) \sim \frac{1.0}{\exp(E/T) \pm 1.0}
\end{equation} 
for the hypothesis of thermalisation to be true. For a specific value of $\tilde{\xi}^{\pm}$, the limits of the above integrals will have the following form:
\begin{equation}
p_{T,max}^2 = (\xi^{\pm}\sqrt{s})^{2} - m^{2} \label{ptmax}
\end{equation}
\begin{equation}
p_{max} = \frac{-\tilde{\xi^{\pm}}\sqrt{s}cos(\theta) + \sqrt{(\tilde{\xi^{\pm}}\sqrt{s})^2 - m^2 sin^2(\theta)}}{sin^2(\theta)}
\label{pmax}
\end{equation}
\begin{equation}
p_{z,max} = \frac{m^2 + p_T^2 - (\tilde{\xi^{\pm}}\sqrt{s})^2}{-2\tilde{\xi^{\pm}}\sqrt{s}}
\label{pzmax}
\end{equation}
Such a definition of the contour of integration encapsulates the closed paraboloidal region in the phase space given by Eq.\eqref{paraboloid}. A successful description of the respective spectra with this construction in \cite{Amaglobeli99, Djobava03, Chkhaidze2006} made it possible to conclude that the hypothesis of thermalisation in the region of the phase space with $\zeta^{\pm} > \tilde{\zeta}^{\pm}$ is true for the $\pi^{\pm}$ (An analysis with $f(E)$ taking the Boltzmannian form ($\exp(-E/T)$) was also made in \cite{Amaglobeli99, Djobava03, Chkhaidze2006} to reach the same conclusions). The analysis was extended to the Au-Au collisions at RHIC energies using the Ultrarelativistic Quantum Molecular Dynamics (UrQMD) model \cite{UrQMD1, UrQMD2} in \cite{nair2021light}. The feasibility of performing the analysis with the ALICE experiment at LHC was studied in \cite{nair2021feasibility} and the $p_T$ dependence of this scheme was explored within the UrQMD model in \cite{nair2021polynomial}. In this paper, we present the light front analysis of simulated pp collisions at $\sqrt{s} = 13$ TeV using PYTHIA-8 event generator \cite{Sj_strand_2006, SJOSTRAND2008852} to investigate the possible existence of thermalised $\pi^{\pm}$, $K^{\pm}$ and $p(\bar{p})$ in the phase space of these collisions. Since there is no QGP-like medium implemented explicitly in PYTHIA 8, the analysis may be considered as a baseline for investigating the QGP formation and thermalisation in the high energy pp collisions at LHC experiments using the light front scheme of analysis.
\section{PYTHIA - 8 : Tuning and trigger} 
The perturbative QCD based PYTHIA 8 (Version- 8303) event generator is used to simulate pp collisions at $\sqrt{s} = 13$ TeV \cite{Sj_strand_2006, SJOSTRAND2008852}. The fragmentation in the PYTHIA event generator is based on the Lund string model \cite{ANDERSSON198331, SJOSTRAND1984469}. The collectivity in pp collisions, linear increase of $J/\psi$ production with multiplicity at the forward rapidities in pp collisions at the LHC are reproduced through the Multi-Parton Interactions (MPI) via the Colour Reconnection (CR) mechanism in PYTHIA 8 \cite{sjostrand2013colour, PhysRevLett.111.042001, ORTIZ201578, 2012165, PhysRevD.97.094002, PhysRevD.81.052012, 20081}. The CR mechanism also modifies the final string fragmentation into hadrons \cite{PhysRevD.92.094010, PhysRevD.97.036010}. To make the spectra of particles closer to the experimentally observed distributions, a tuning is implemented for the generation of the event with Monash 2013 tune \cite{Skands_2014}. The inelastic, non-diffractive component of the total cross-section is implemented for all soft QCD processes and the MPI based scheme of CR is implemented and all the resonances are allowed to decay. The trigger criterion is such that only those events with at least 5 charged particles in the pseudorapidity range of $|\eta| < 0.8$ are considered for the analysis. The charged particle multiplicities have been chosen in the acceptance of V0 detectors in ALICE at the LHC \cite{V02013} with pseudo-rapidity coverage of V0A ($2.8 < \eta < 5.1$) and V0C ($-3.7 < \eta < -1.7$). Those events with $50-140$ charged particles in this range of acceptance is taken as the top 10\% events \cite{doi:10.1142/S0217751X14300440, Khuntia_2021}. Hundred thousand simulated pp events which survived the trigger criterion with the tuning is considered for further analysis. The schemes of analysis implemented and the results obtained from them are presented in the next section.
\section{Schemes of analysis and results} 
Two schemes of analysis are implemented separately over the inclusively produced $\pi^{\pm}$, $K^{\pm}$ and $p(\bar{p})$ in the simulated collisions. The steps involved in those two schemes and the results of the analysis are presented in this section.
\subsection*{Scheme - 1} 
\begin{figure*}[hbt!] 
\begin{subfigure}{0.65\textwidth}
\includegraphics[width=\linewidth]{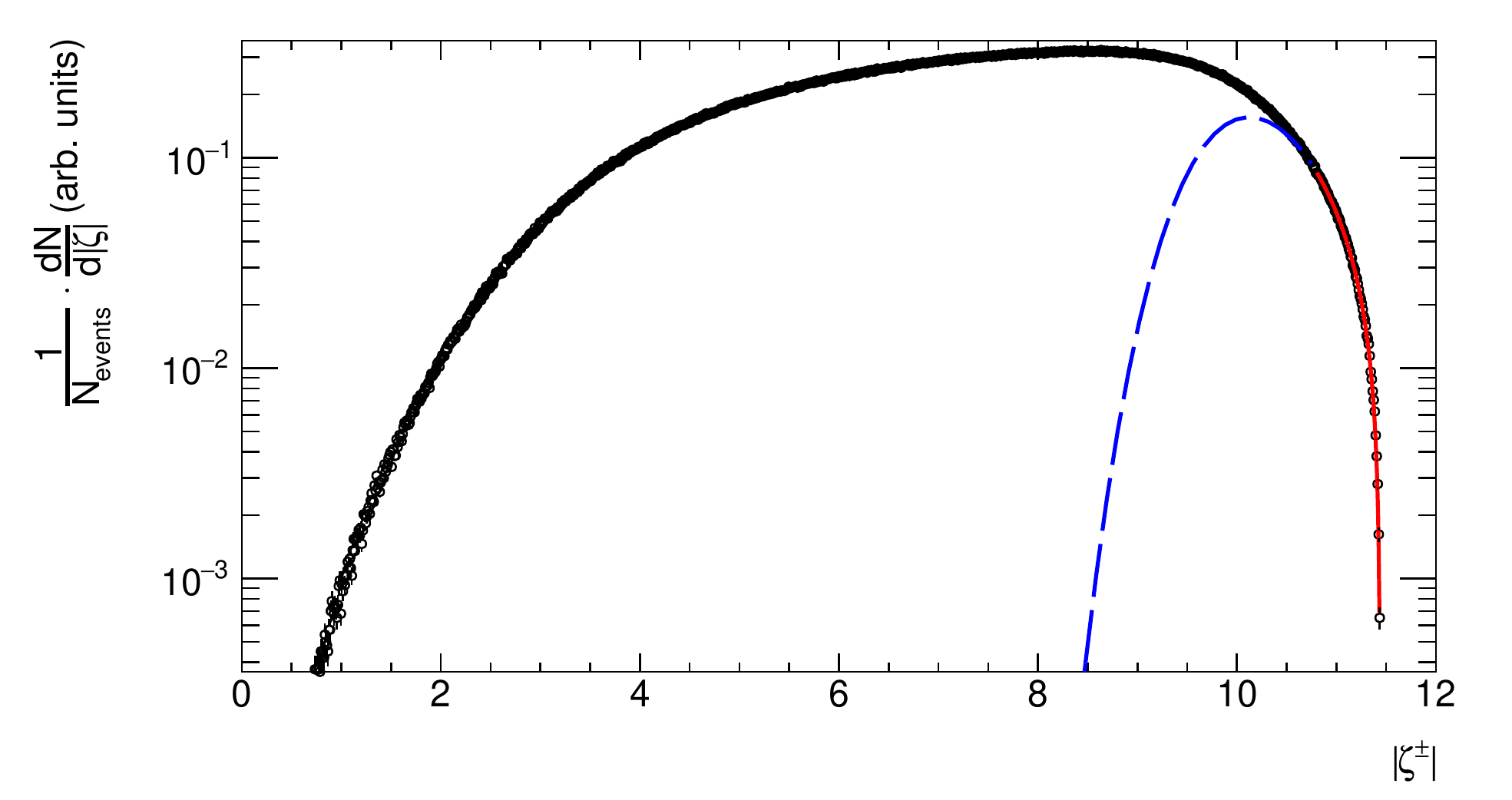}
\caption{$|\zeta^{\pm}|$ distribution of $\pi^{\pm}$; $\tilde{\zeta}^{\pm} = 10.80$ }\label{ZetaPion}
\end{subfigure}\\
\begin{subfigure}{0.65\textwidth}
\includegraphics[width=\linewidth]{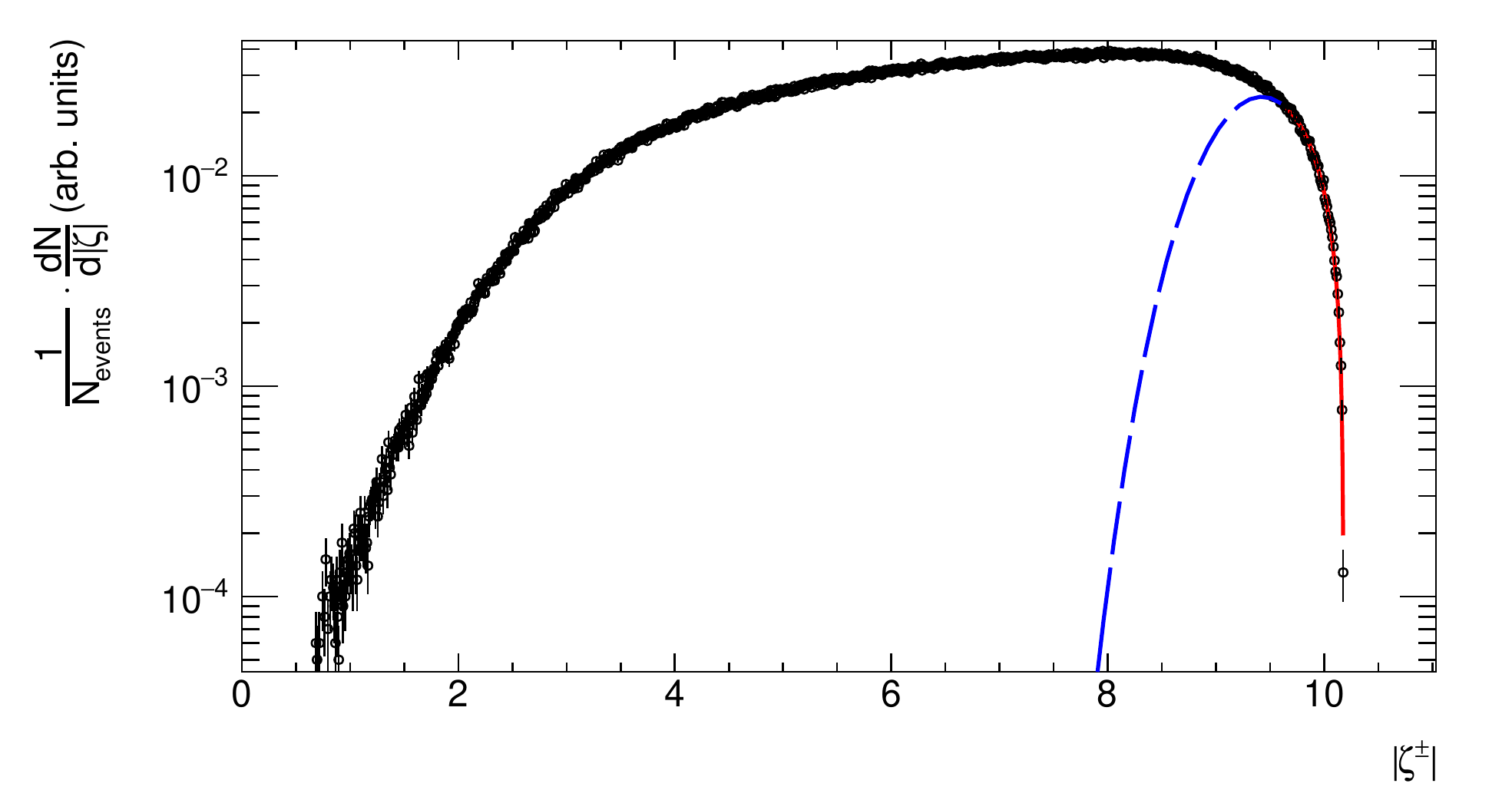}
\caption{$|\zeta^{\pm}|$ distribution of $K^{\pm}$; $\tilde{\zeta}^{\pm} = 9.65$}
\label{ZetaKaon}
\end{subfigure}\\
\begin{subfigure}{0.65\textwidth}
\includegraphics[width=\linewidth]{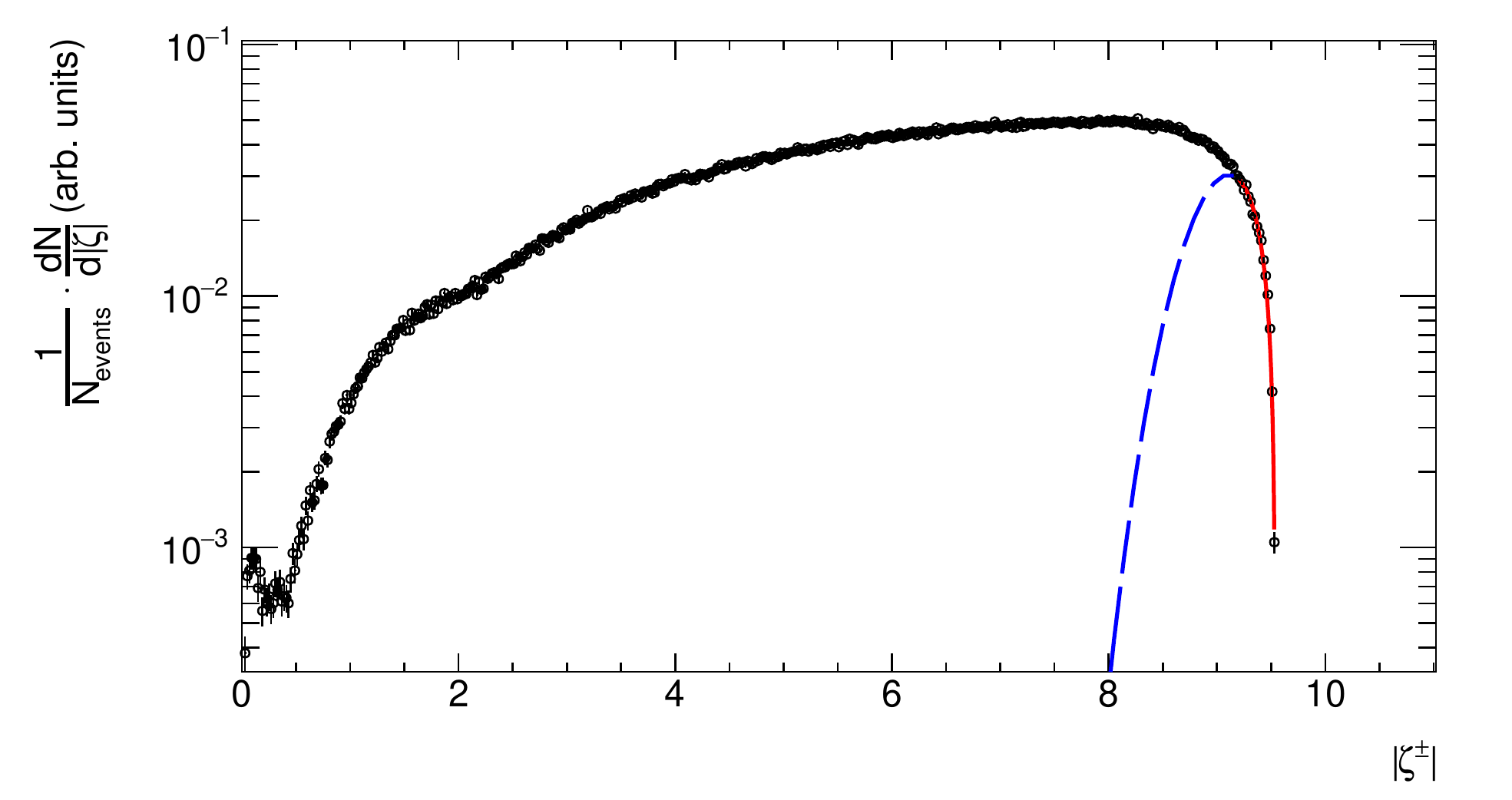}
\caption{$|\zeta^{\pm}|$ distribution of $p(\bar{p})$; $\tilde{\zeta}^{\pm} = 9.20$}
\label{ZetaProton}
\end{subfigure}
\caption{$|\zeta^{\pm}|$ distribution of particles fitted with Eq.\eqref{ZetaInt}. The solid curve is the result of the fit up to $\zeta_c$ and the dashed curve is the extrapolation of the fit beyond the corresponding $\zeta_c$ for each species of particles.} \label{PYTHIAZetaPlots}
\end{figure*}

\begin{figure*}[hbt!] 
\begin{subfigure}{0.65\textwidth}
\includegraphics[width=\linewidth]{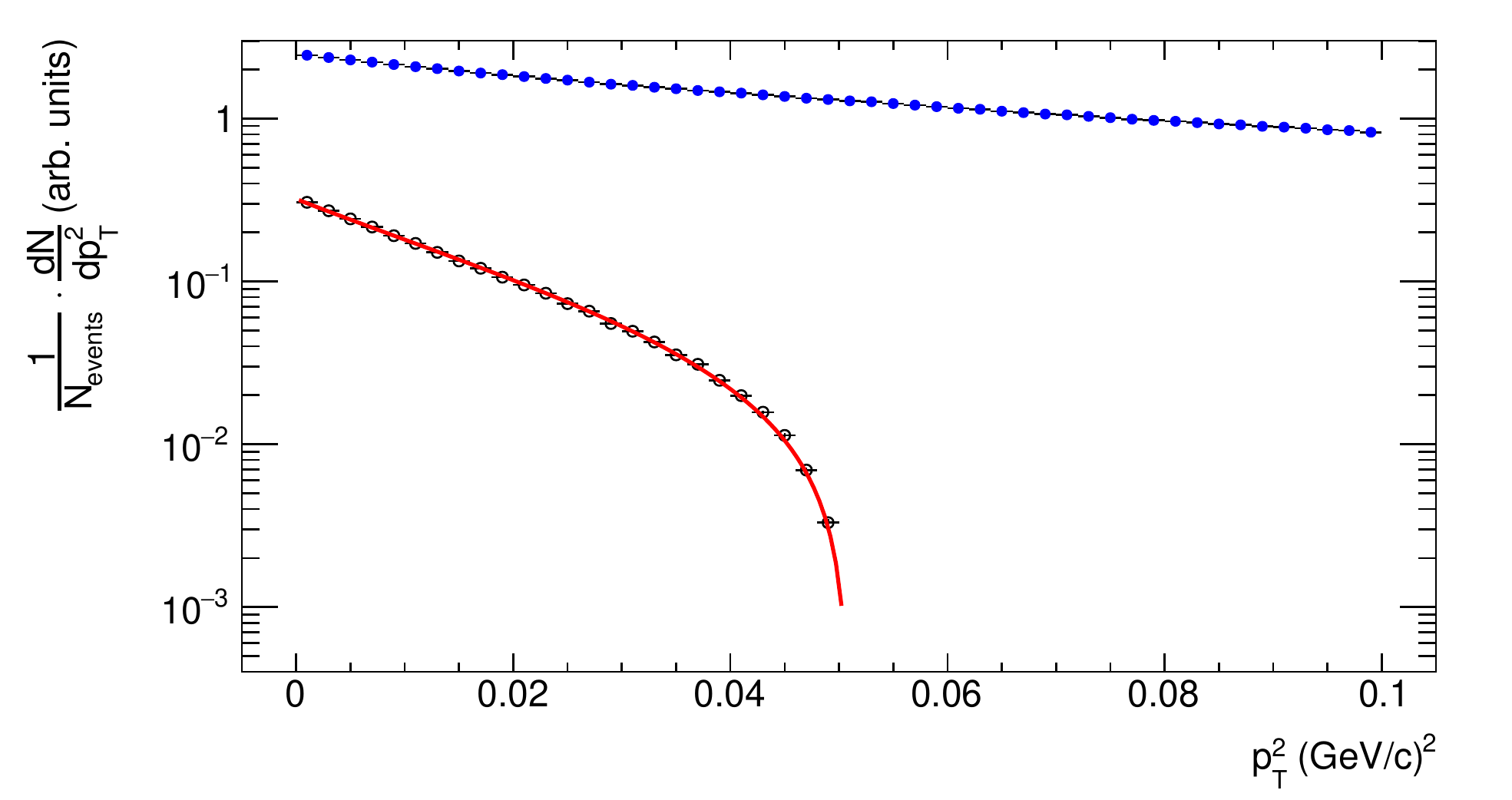}
\caption{$p_{T}^{2}$ distribution of $\pi^{\pm}$; $\tilde{\zeta}^{\pm} = 10.80$ }\label{PtSqPion}
\end{subfigure}\\
\begin{subfigure}{0.65\textwidth}
\includegraphics[width=\linewidth]{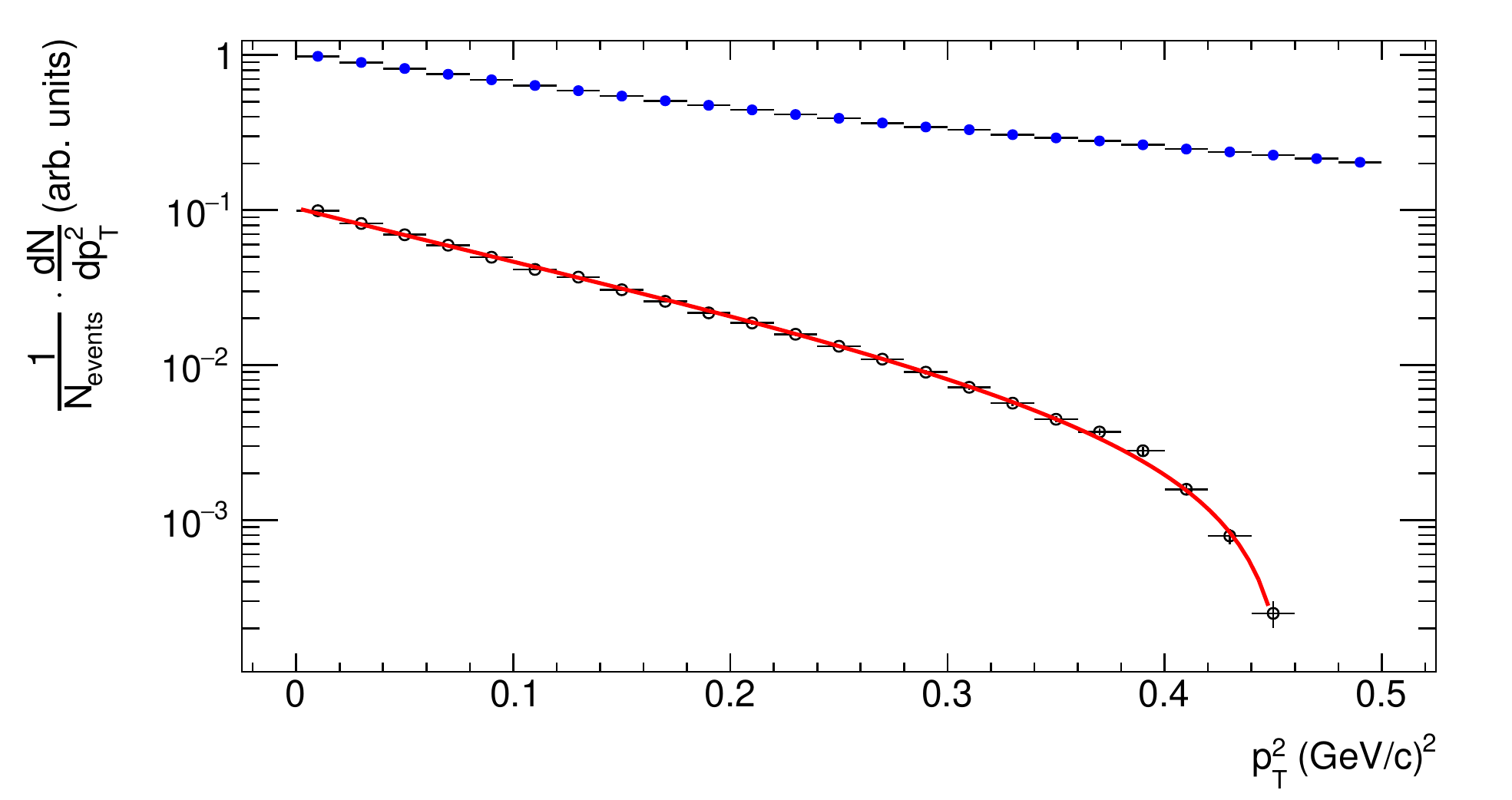}
\caption{$p_{T}^{2}$ distribution of $K^{\pm}$; $\tilde{\zeta}^{\pm} = 9.65$}
\label{PtSqKaon}
\end{subfigure}\\
\begin{subfigure}{0.65\textwidth}
\includegraphics[width=\linewidth]{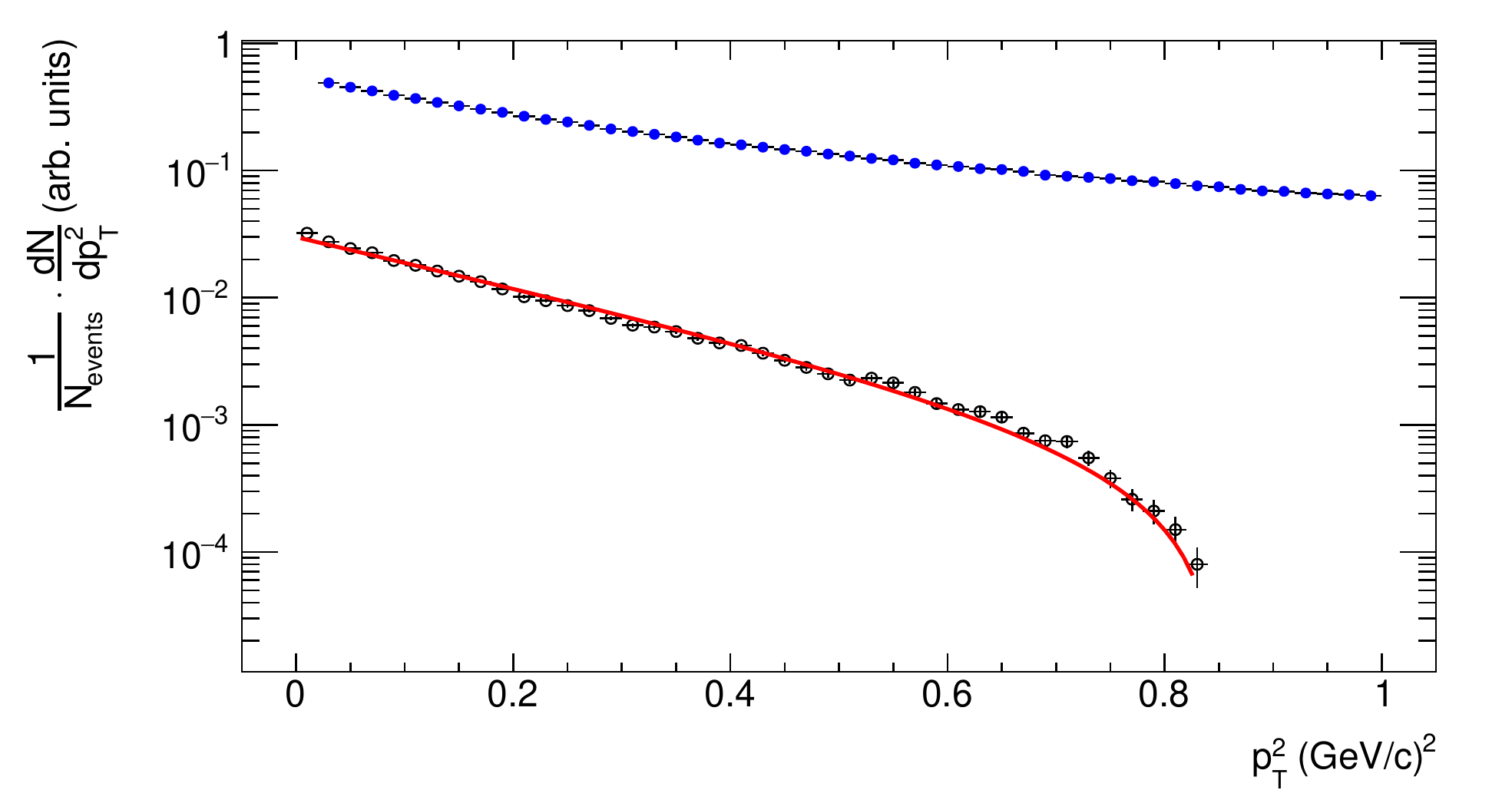}
\caption{$p_{T}^{2}$ distribution of $p(\bar{p})$; $\tilde{\zeta}^{\pm} = 9.20$}
\label{PtSqProton}
\end{subfigure}
\caption{Marked in open circles are the $p_{T}^{2}$ distribution of particles having $\zeta_c > \zeta$ fitted with Eq.\eqref{PtSqInt}. The solid red curves are the result of the fit. Marked in blue are the $p_{T}^{2}$ distribution of the particles with $\zeta_c < \zeta$} \label{PYTHIAPtSqPlots}
\end{figure*}

\begin{figure*}[hbt!] 
\begin{subfigure}{0.65\textwidth}
\includegraphics[width=\linewidth]{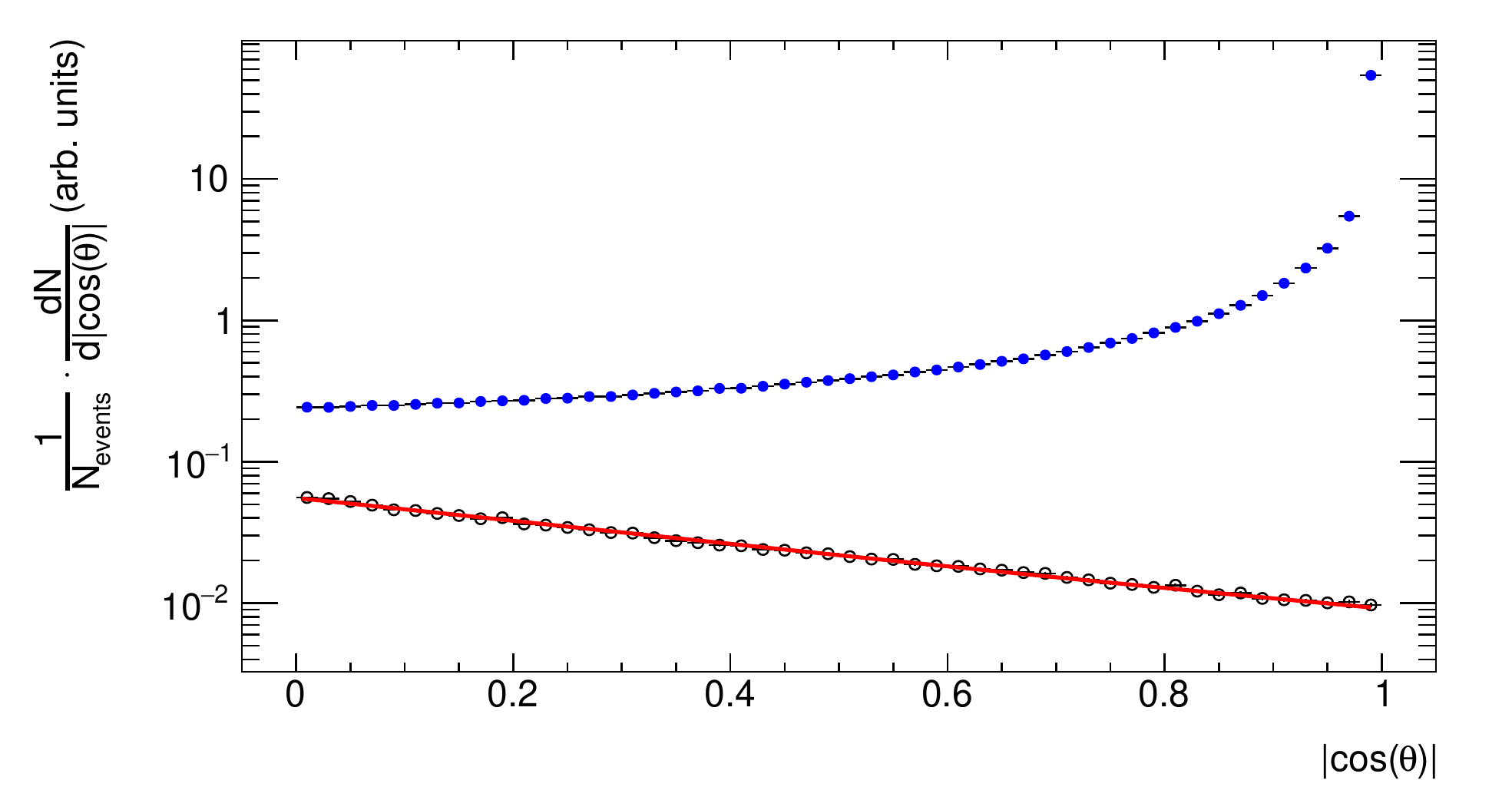}
\caption{$|cos(\theta)|$ distribution of $\pi^{\pm}$; $\tilde{\zeta}^{\pm} = 10.80$ }\label{CosPion}
\end{subfigure}\\
\begin{subfigure}{0.65\textwidth}
\includegraphics[width=\linewidth]{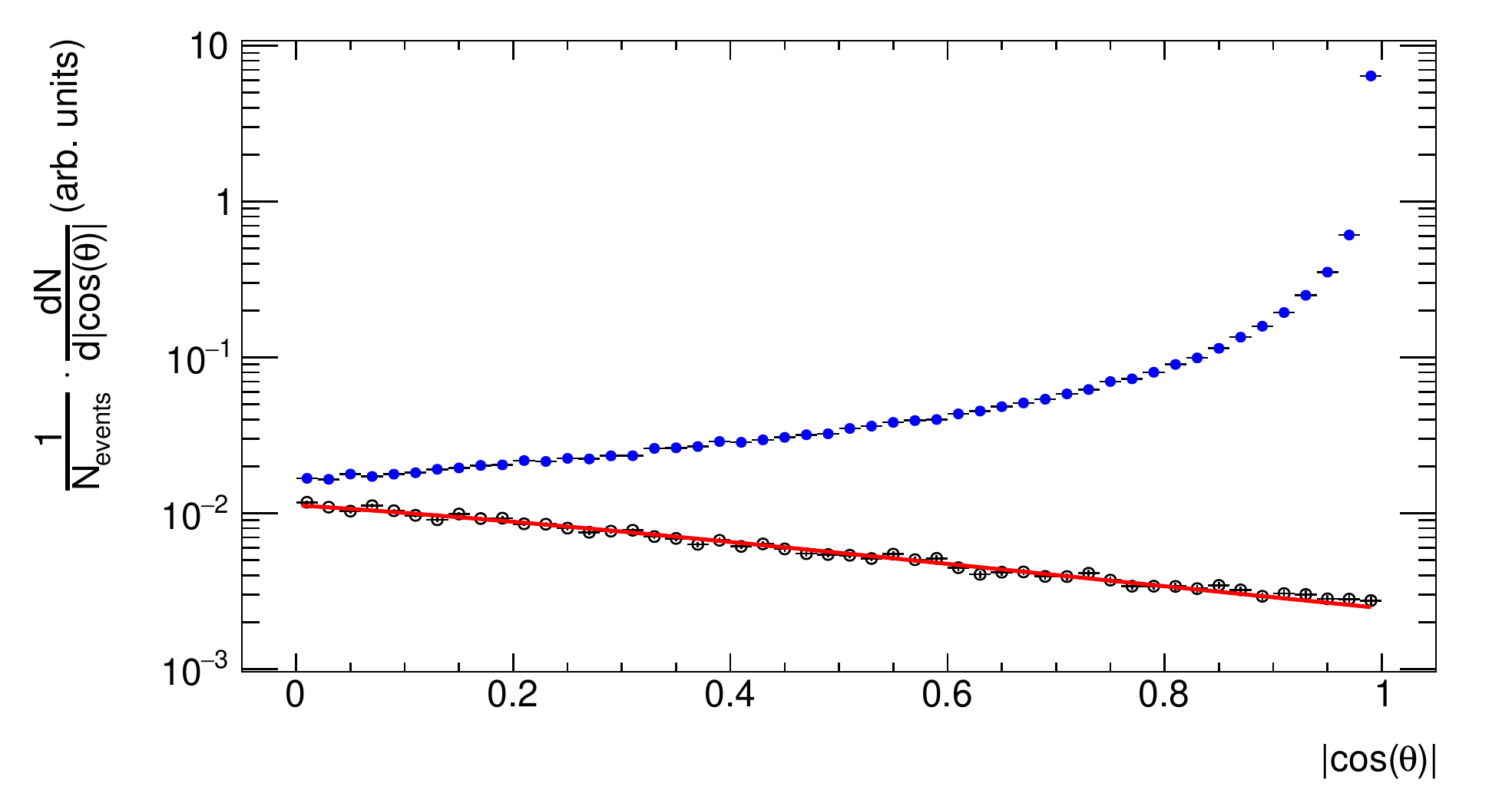}
\caption{$|cos(\theta)|$ distribution of $K^{\pm}$; $\tilde{\zeta}^{\pm} = 9.65$}
\label{CosKaon}
\end{subfigure}\\
\begin{subfigure}{0.65\textwidth}
\includegraphics[width=\linewidth]{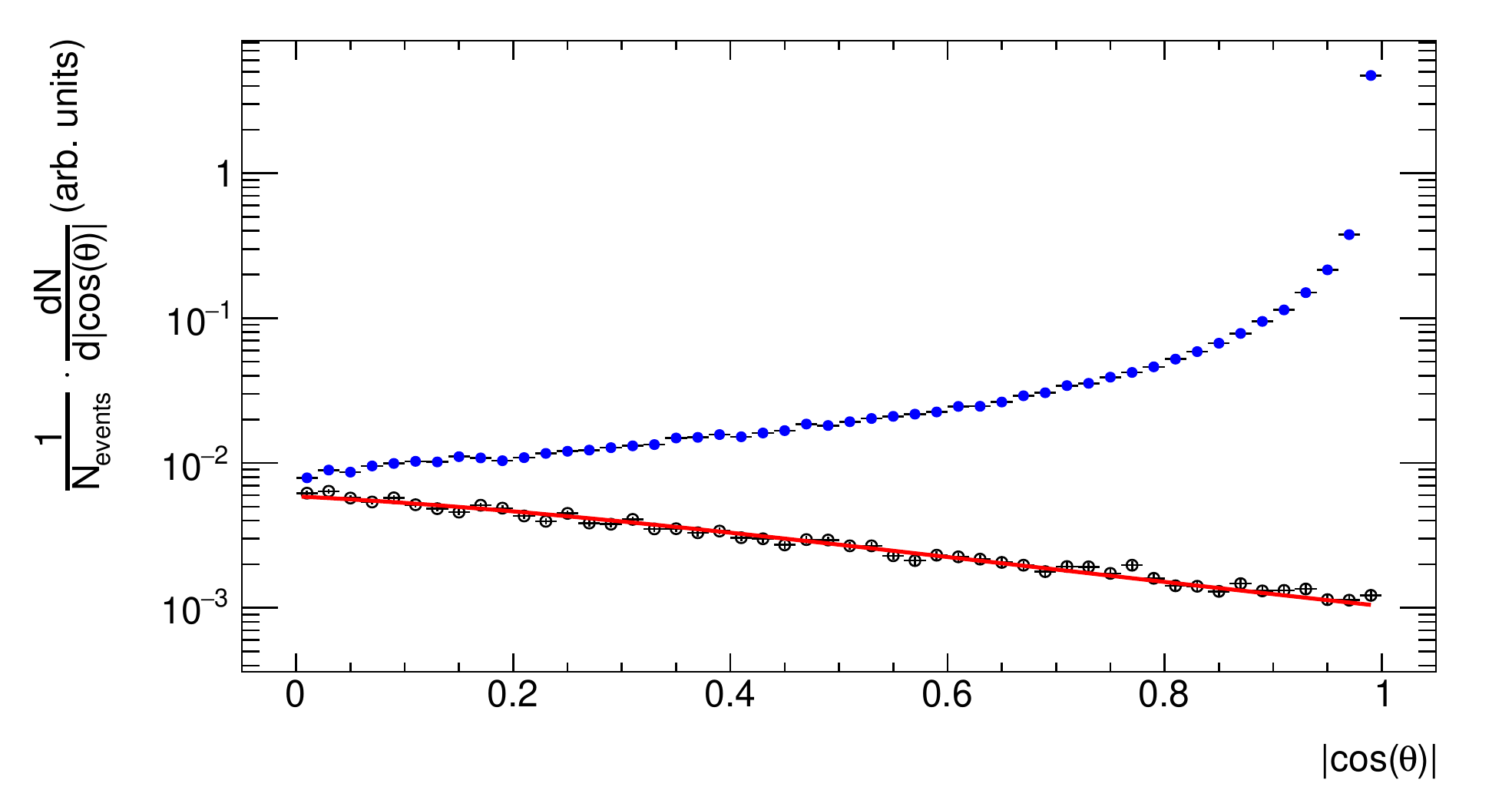}
\caption{$|cos(\theta)|$ distribution of $p(\bar{p})$; $\tilde{\zeta}^{\pm} = 9.20$}
\label{CosProton}
\end{subfigure}
\caption{Marked in open circles are the $|cos(\theta)|$ distribution of particles having $\zeta_c > \zeta$ fitted with Eq.\eqref{CosInt}. The solid red curves are the result of the fit. Marked in blue are the $|cos(\theta)|$ distribution of the particles with $\zeta_c < \zeta$} \label{PYTHIACosPlots}
\end{figure*}

In scheme-1, we closely follow the analysis methodology in \cite{Amaglobeli99}, \cite{nair2021light} etc to search for a certain specific light front surface in the phase space of the particles. The steps involved in the analysis are enlisted below:
\begin{enumerate}

\item The inclusively produced $\pi^{\pm}$, $K^{\pm}$ and $p(\bar{p})$ from the simulated events with PYTHIA 8 are selected and a distribution of $|\zeta^{\pm}|$ is made for each of them. For $\zeta^{\pm}$ and $cos(\theta)$ we take the absolute values to fill the distributions owing to the symmetry of these distributions for the two hemispheres in pp collisions.
  
\item The $|\zeta^{\pm}|$ distributions are fitted with Eq. \eqref{ZetaInt}. The lowest value of $\zeta$ down to which it can be done successfully is taken as $|\tilde{\zeta^{\pm}}|$. We use the $\chi^2$ minimisation method for fitting as implemented in the ROOT (version 6) software package \cite{BRUN199781}. For a successful fit the relation
\begin{equation}\label{ChiSq} 
\frac{\chi^2}{n.d.f} \sim 1.0
\end{equation}
will hold true where $n.d.f$ stands for the number of degrees of freedom. 
  
\item The particles are then divided into two groups based on the value of the $|\tilde{\zeta^{\pm}}|$. The $|cos(\theta)|$ distribution of particles with $|\zeta^{\pm}| > |\tilde{\zeta^{\pm}}|$ is made and is fitted with Eq. \eqref{CosInt}. The $|\tilde{\zeta^{\pm}}|$ in the limit of the integral denoted by $p_{max}$ as in Eq.\eqref{pmax} is the value of $|\tilde{\zeta^{\pm}}|$ obtained from fitting the $|\zeta^{\pm}|$ distribution in the previous step. 
  
\item The $p_{T}^{2}$ distribution of particles with $|\zeta^{\pm}| > |\tilde{\zeta^{\pm}}|$ is made and fitted with Eq. \eqref{PtSqInt} using the same value $|\tilde{\zeta^{\pm}}|$ for calculating $p_{z,max}$ as in Eq.\eqref{pzmax}. In all these three cases of the fitting the form of $f(E)$ is taken as shown in Eq.\eqref{eqboltz}. 
  
\item If the three fits are successful, then $|\tilde{\zeta^{\pm}}|$ is taken as the final value $\zeta_c$ of the light front variable. If the fits do not follow Eq. \eqref{ChiSq}, we repeat the steps from 2 to 4 with a larger value of $|\tilde{\zeta^{\pm}}|$ until the three fits are successful or the value of $|\zeta^{\pm}|$ can no longer be increased.

\item If we can find a $\zeta_c$ from the above procedure for a specific species of particle, we say that the light front analysis could select a thermalised group of particles of that kind. Instead, if the fitting cannot be done successfully for any values of the $|\tilde{\zeta^{\pm}}|$, then we say that the light front analysis fails to find a thermalised group for the corresponding species.
\end{enumerate}
The above strategy was implemented for the generated data. For the three species of particles considered in our analysis, we could find a $\zeta_c$ and hence a group of thermalised particles. The fitted $|\zeta^{\pm}|$, $p_{T}^{2}$ and $|cos(\theta)|$ distributions are shown in FIG.\ref{PYTHIAZetaPlots}, FIG.\ref{PYTHIAPtSqPlots} and FIG.\ref{PYTHIACosPlots} respectively. The temperatures obtained from the analysis are summarised in TABLE.\ref{PYTHIATempratureTable}.
\begin{table*}[hbt!]
\begin{center}
\begin{tabular}{lllllllll} 
\hline\hline \noalign{\smallskip}
Species & $\;\;\;\zeta_c\;\;\;$ & $T_{\zeta}$(MeV)& $\chi^2/ndf$& $T_{p_T^2}$(MeV) & $\chi^2/ndf$ & $T_{cos(\theta)}$ (MeV)& $\chi^2/ndf$\\ 
\noalign{\smallskip}\hline\noalign{\smallskip} 
$\pi^{\pm}$ & 10.80 & 118 $\pm$ 3 & 23/62 & 96 $\pm$ 6 & 2/23 & 82 $\pm$ 2 & 41/48 \\\noalign{\smallskip}
$K^{\pm}$  & 9.65 & 195 $\pm$ 5   & 46/51 & 154 $\pm$ 7 & 2/21 & 123 $\pm$ 3 & 54/48 \\\noalign{\smallskip}
$p(\bar{p})$ & 9.20 & 188 $\pm$ 9 & 12/15 & 139 $\pm$ 3 & 42/39 & 115 $\pm$ 4 & 59/48 \\\noalign{\smallskip}
\noalign{\smallskip}\hline\hline
\end{tabular}
\end{center}
\caption{Temperatures obtained from the naive light front analysis of the hadrons in PYTHIA}
\label{PYTHIATempratureTable}
\end{table*}
\begin{figure*}[hbt!]
\centering
\includegraphics[width=\textwidth]{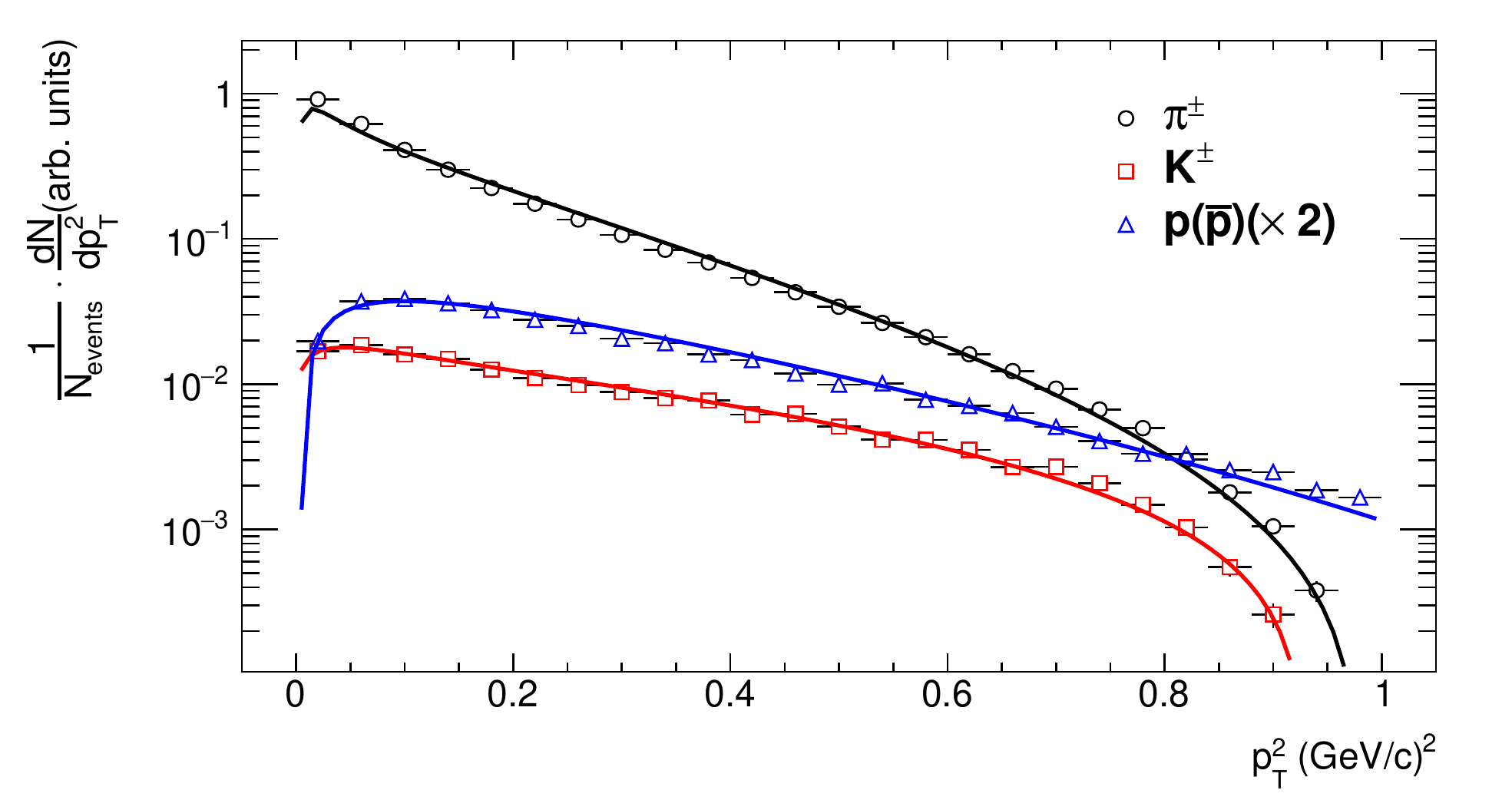}
\caption{$p_T^2$ distributions of hadrons with $|\zeta^{\pm}| > \zeta_{c}(p_T,m)$ fitted with Eq.\eqref{PtSqInt} as per scheme-2 with the upper limit of the integration given by Eq.\eqref{pzmaxPoly}. Solid curves are the result of the fits.}
\label{ptsqmesons}
\end{figure*}
\subsection*{Scheme - 2}
\begin{table*}[hbt!]
\begin{center}
\setlength{\tabcolsep}{10pt} 
\renewcommand{\arraystretch}{1} 
\begin{tabular} {llllllll}
\hline\hline \noalign{\smallskip}
Species    &$a_{0}$& $a_{1}$& $a_{2}$ & $a_{3}$  & T(MeV) &$\chi^2/n.d.f$ \\
\noalign{\smallskip}\hline\noalign{\smallskip}
$\pi^{\pm}$  & 11.542 & -4.82366 & 4.30804 & -1.5625 & 160$\pm$4 & 10/23 \\\noalign{\smallskip}
$K^{\pm}$   & 10.1648 & -0.331889 & -1.49097 & 1.0416 & 176$\pm$4 & 15/21 \\ \noalign{\smallskip}
$p(\bar{p})$ & 9.60808 & -1.19122 & 1.22768 & -0.520833 & 162$\pm$4 & 25/23 \\ \noalign{\smallskip}
\noalign{\smallskip}\hline\hline
\end{tabular}
\end{center}
\caption{Results of the fit of $p_T^2$ distributions of particles with their $|\zeta^{\pm}| > \zeta_{c}(p_T,m)$ using the relation in Eq.\eqref{PtSqInt}}
\label{TableIntegral}
\end{table*}  
In scheme -2, we closely follow the analysis methodology in \cite{nair2021polynomial} to investigate whether we can establish a polynomial in $p_T$ for the value of $\zeta_{c}$ of the particles in the pseudorapidity range of $|\eta| < 0.8$. Instead of looking for a specific $\zeta_{c}$ for all of the particles (of a specific species) in the phase space like in scheme-1, here we try to establish two paraboloids in the phase space for a specific particle: one corresponding to the $\zeta$ value of the particle and the second one corresponding to the $\zeta_c(p_T)$ of the particle. The steps involved in this analysis are as follows:
\begin{enumerate}
\item The inclusively produced $\pi^{\pm}$, $K^{\pm}$ and $p(\bar{p})$ from PYTHIA 8 simulated events are selected and a distribution of $|\zeta^{\pm}|$ is made for each of them for different bins in $p_T$ in the mid-rapidity region ($|\eta| < 0.8$).

\item The $|\zeta^{\pm}|$ distribution for each $p_T$ bin are then fitted with Eq. \eqref{ZetaInt} taking into account the kinematic constraints. The lowest value of $\zeta$ to which it can be done successfully is taken as $|\tilde{\zeta^{\pm}}|$ for the corresponding $p_T$ bin. 

\item The $|cos(\theta)|$ distribution of particles with $|\zeta^{\pm}| > |\tilde{\zeta^{\pm}}|$ is made for each $p_T$ bin and is fitted with Eq. \eqref{CosInt}. Similarly the $p_{T}^{2}$ distribution of particles with $|\zeta^{\pm}| > |\tilde{\zeta^{\pm}}|$ is made and fitted with Eq. \eqref{PtSqInt} using the same value $|\tilde{\zeta^{\pm}}|$ for the $p_T$ bin

\item If the three fits are successful, then $|\tilde{\zeta^{\pm}}|$ is taken as the final value $\zeta_c$ for the corresponding $p_T$ bin. If the fits do not follow Eq. \eqref{ChiSq}, we repeat the steps from 1 to 3 with a larger value of $|\tilde{\zeta^{\pm}}|$ until the three fits are successful or the value of $|\zeta^{\pm}|$ can no longer be increased kinematically.

\item From the above mentioned procedure, an approximate polynomial relationship of the form 
\begin{equation}\label{polynomial}
\zeta_{c}(p_T,m) = a_{0}+ a_{1}p_T + a_{2}p_T^2 + a_{3}p_T^3
\end{equation}
is constructed. Thus for a specific particle, we have a $\zeta$ from Eq.\eqref{EqnZeta} and a $\zeta_{c}(p_T,m)$ from Eq.\eqref{polynomial}.

\item In the final step, we make the $p_T^2$ distributions of those particles with their $|\zeta^{\pm}| > \zeta_{c}(p_T,m)$ and try to fit it with Eq.\eqref{PtSqInt}. The upper limit of the integral in Eq.\eqref{PtSqInt} in this scheme given by 
\begin{equation}
p_{z,max} = \frac{m^2 + p_{T}^2 - (\sqrt{s}e^{ -\zeta_{c}(p_T,m)})^2}{-2\sqrt{s}e^{ -\zeta_{c}(p_T,m)}}
\label{pzmaxPoly}
\end{equation}
\end{enumerate}
For the three species of particles we considered in the analysis, a polynomial relation of the form as in Eq.\eqref{polynomial} could be set up between $\zeta_{c}$ and $p_T$ of the particle with the coefficients as given in TABLE \ref{TableIntegral} in the mid-rapidity region. With these values of the coefficients, the $p_T^2$ distributions of the particles with $|\zeta^{\pm}| > \zeta_{c}(p_T,m)$ could be fitted with Eq.\eqref{PtSqInt} while the limits are given by Eq.\eqref{pzmaxPoly}. The results of the fits are shown in the FIG. \ref{ptsqmesons}. Thus it follows from our analysis that a selection can be made for the particles regarding their thermal equilibration based on the relative position of the two paraboloids defined by $\zeta$ and $\zeta_c$ of the particle. Moreover, it can be used as a tool to test the presence of thermalisation in high energy collisions. The interpretation of the values of the obtained temperatures requires further studies.

\section{Conclusions and Discussions} 
The proton-proton collisions at high energies are considered to be a benchmark for comparison of the results from the ultrarelativistic heavy-ion collisions where a deconfined state of quarks and gluons is supposedly formed. The experimentally observed signatures of QGP-like medium in the high multiplicity pp collisions create concerns over the benchmarking of these collisions for interpreting the results from ultrarelativistic heavy-ion collisions. In the heavy-ion physics community, the describability of the experimentally observed collective flow coefficients with hydrodynamical models which assume a local thermal equilibration is generally interpreted as the evidence of the presence of a QGP-like medium. Hence understanding the dynamics of the pp collisions and investigating the possible formation of quark-gluon plasma (QGP) like medium in such collisions are of utmost importance.\par We approach the scenario from the perspective of a so-called light front analysis. Analysis of a similar kind with the low energy collisions could establish a constant paraboloidal surface in the phase space of inclusively produced charged pions that could select a group of particles following the thermal Boltzmann distribution \cite{Amaglobeli99, Djobava03, Chkhaidze2006}. We extend this analysis to the pp collisions at high energies and propose it as a tool to investigate the thermalisation in pp collisions at LHC. Two schemes of light front analysis are performed over the inclusively produced $\pi^{\pm}$, $K^{\pm}$ and $p(\bar{p})$ in the central pp collisions simulated using the PYTHIA 8 event generator at $\sqrt{s} = 13$ TeV. The main outcomes of the analysis are summarised below:
\begin{enumerate}

\item A specific paraboloidal surface could be established in the phase of the inclusively produced $\pi^{\pm}$, $K^{\pm}$ and $p(\bar{p})$ in the central pp collisions simulated with PYTHIA 8, which can select a group of particles following the Bose-Einstein (for $\pi^{\pm}$ and $K^{\pm}$) or Fermi-Dirac (for $p(\bar{p})$) statistics.

\item The $p_T$ dependence of light front analysis is explored in the mid rapidity region for the inclusively produced $\pi^{\pm}$, $K^{\pm}$ and $p(\bar{p})$. We could establish a certain light front value value $\zeta_{c}(p_T)$ of the light front variable for each of the particle species as a polynomial in its $p_T$, which can distinguish a group of $\pi^{\pm}$ and $K^{\pm}$ following Bose-Einstein  and $p(\bar{p})$ Fermi-Dirac statistics, upon comparison with the numerical value of the actual light front value $|\zeta^{\pm}|$ of the particle.

\item The temperatures we obtained from the two schemes are different for all three species. Similar to the case with the older studies implementing the light front analysis as in scheme-1, a convergence of temperatures from fitting the $|\zeta^{\pm}|$ and $p_T^2$ $|cos(\theta)|$ distributions is not observed in scheme-1 for the constant values of the light front variable we selected. 

\item The PYTHIA 8 event generator does not explicitly incorporate a thermal, deconfined medium in it. Hence the results we obtained from the analysis may be treated as a baseline for the experimental analysis towards the search of QGP-like medium in the central pp collisions at LHC.\\

\end{enumerate}
To conclude, the light front analysis is proposed for studying the formation of the deconfined medium in high multiplicity pp collisions at LHC experiments. The possibility of the formation of a deconfined state of partons may be studied by implementing the analysis over the experimental data and comparing its outcomes with the results of our analysis. Such a study might give insights into the origin of collective effects observed in high energy pp collisions as well. It should be noted that the exact reason for the light front analysis to work as it does is not known. No acceptable theoretical explanation exist about the characteristics of a constant light front variable or a polynomial in $p_T$ giving a specific light front value that select a thermalised group of particles. \par The curious nature of this observation thus exists in its original fashion from the very first studies on this subject. One may claim that the hydrodynamical studies do tell us about the thermalisation of hadrons in nuclear collisions at sufficiently low $p_T$. Experimental exploration of the status of $\zeta_{c}(p_T)$ as polynomial that can select a thermalised group of particles at very high values of $p_T$ might provide more insights in this respect. However, irrespective of its origin being naive or profound, the approach as demonstrated in this paper can be useful in handling several scenarios in high energy hadron-hadron and nucleus-nucleus interactions.

\bibliographystyle{apsrev4-2}
\bibliography{article}

\begin{thebibliography}{37}%
\makeatletter
\providecommand \@ifxundefined [1]{%
 \@ifx{#1\undefined}
}%
\providecommand \@ifnum [1]{%
 \ifnum #1\expandafter \@firstoftwo
 \else \expandafter \@secondoftwo
 \fi
}%
\providecommand \@ifx [1]{%
 \ifx #1\expandafter \@firstoftwo
 \else \expandafter \@secondoftwo
 \fi
}%
\providecommand \natexlab [1]{#1}%
\providecommand \enquote  [1]{``#1''}%
\providecommand \bibnamefont  [1]{#1}%
\providecommand \bibfnamefont [1]{#1}%
\providecommand \citenamefont [1]{#1}%
\providecommand \href@noop [0]{\@secondoftwo}%
\providecommand \href [0]{\begingroup \@sanitize@url \@href}%
\providecommand \@href[1]{\@@startlink{#1}\@@href}%
\providecommand \@@href[1]{\endgroup#1\@@endlink}%
\providecommand \@sanitize@url [0]{\catcode `\\12\catcode `\$12\catcode
  `\&12\catcode `\#12\catcode `\^12\catcode `\_12\catcode `\%12\relax}%
\providecommand \@@startlink[1]{}%
\providecommand \@@endlink[0]{}%
\providecommand \url  [0]{\begingroup\@sanitize@url \@url }%
\providecommand \@url [1]{\endgroup\@href {#1}{\urlprefix }}%
\providecommand \urlprefix  [0]{URL }%
\providecommand \Eprint [0]{\href }%
\providecommand \doibase [0]{https://doi.org/}%
\providecommand \selectlanguage [0]{\@gobble}%
\providecommand \bibinfo  [0]{\@secondoftwo}%
\providecommand \bibfield  [0]{\@secondoftwo}%
\providecommand \translation [1]{[#1]}%
\providecommand \BibitemOpen [0]{}%
\providecommand \bibitemStop [0]{}%
\providecommand \bibitemNoStop [0]{.\EOS\space}%
\providecommand \EOS [0]{\spacefactor3000\relax}%
\providecommand \BibitemShut  [1]{\csname bibitem#1\endcsname}%
\let\auto@bib@innerbib\@empty
\bibitem [{\citenamefont {Koch}\ \emph {et~al.}(2017)\citenamefont {Koch},
  \citenamefont {Müller},\ and\ \citenamefont
  {Rafelski}}]{doi:10.1142/S0217751X17300241}%
  \BibitemOpen
  \bibfield  {author} {\bibinfo {author} {\bibfnamefont {P.}~\bibnamefont
  {Koch}}, \bibinfo {author} {\bibfnamefont {B.}~\bibnamefont {Müller}},\ and\
  \bibinfo {author} {\bibfnamefont {J.}~\bibnamefont {Rafelski}},\ }\href
  {https://doi.org/10.1142/S0217751X17300241} {\bibfield  {journal} {\bibinfo
  {journal} {International Journal of Modern Physics A}\ }\textbf {\bibinfo
  {volume} {32}},\ \bibinfo {pages} {1730024} (\bibinfo {year}
  {2017})}\BibitemShut {NoStop}%
\bibitem [{\citenamefont {Aamodt}\ \emph {et~al.}(2008)\citenamefont {Aamodt}
  \emph {et~al.}}]{Collaboration_2008}%
  \BibitemOpen
  \bibfield  {author} {\bibinfo {author} {\bibfnamefont {K.}~\bibnamefont
  {Aamodt}} \emph {et~al.} (\bibinfo {collaboration} {ALICE Collaboration}),\
  }\href {https://doi.org/10.1088/1748-0221/3/08/s08002} {\bibfield  {journal}
  {\bibinfo  {journal} {J. Instrum}\ }\textbf {\bibinfo {volume} {3}},\
  \bibinfo {pages} {S08002} (\bibinfo {year} {2008})}\BibitemShut {NoStop}%
\bibitem [{\citenamefont {Adam}\ \emph {et~al.}(2017)\citenamefont {Adam} \emph
  {et~al.}}]{Adam2017}%
  \BibitemOpen
  \bibfield  {author} {\bibinfo {author} {\bibfnamefont {J.}~\bibnamefont
  {Adam}} \emph {et~al.} (\bibinfo {collaboration} {ALICE Collaboration}),\
  }\href {https://doi.org/10.1038/nphys4111} {\bibfield  {journal} {\bibinfo
  {journal} {Nature Physics}\ }\textbf {\bibinfo {volume} {13}},\ \bibinfo
  {pages} {535} (\bibinfo {year} {2017})}\BibitemShut {NoStop}%
\bibitem [{\citenamefont {Khachatryan}\ \emph {et~al.}(2017)\citenamefont
  {Khachatryan} \emph {et~al.}}]{2017193}%
  \BibitemOpen
  \bibfield  {author} {\bibinfo {author} {\bibfnamefont {V.}~\bibnamefont
  {Khachatryan}} \emph {et~al.} (\bibinfo {collaboration} {CMS
  Collaboration}),\ }\href
  {https://doi.org/https://doi.org/10.1016/j.physletb.2016.12.009} {\bibfield
  {journal} {\bibinfo  {journal} {Physics Letters B}\ }\textbf {\bibinfo
  {volume} {765}},\ \bibinfo {pages} {193} (\bibinfo {year}
  {2017})}\BibitemShut {NoStop}%
\bibitem [{\citenamefont {Bzdak}\ and\ \citenamefont
  {Ma}(2014)}]{PhysRevLett.113.252301}%
  \BibitemOpen
  \bibfield  {author} {\bibinfo {author} {\bibfnamefont {A.}~\bibnamefont
  {Bzdak}}\ and\ \bibinfo {author} {\bibfnamefont {G.-L.}\ \bibnamefont {Ma}},\
  }\href {https://doi.org/10.1103/PhysRevLett.113.252301} {\bibfield  {journal}
  {\bibinfo  {journal} {Phys. Rev. Lett.}\ }\textbf {\bibinfo {volume} {113}},\
  \bibinfo {pages} {252301} (\bibinfo {year} {2014})}\BibitemShut {NoStop}%
\bibitem [{\citenamefont {Bo\ifmmode~\dot{z}\else \.{z}\fi{}ek}\ \emph
  {et~al.}(2013)\citenamefont {Bo\ifmmode~\dot{z}\else \.{z}\fi{}ek},
  \citenamefont {Broniowski},\ and\ \citenamefont
  {Torrieri}}]{PhysRevLett.111.172303}%
  \BibitemOpen
  \bibfield  {author} {\bibinfo {author} {\bibfnamefont {P.}~\bibnamefont
  {Bo\ifmmode~\dot{z}\else \.{z}\fi{}ek}}, \bibinfo {author} {\bibfnamefont
  {W.}~\bibnamefont {Broniowski}},\ and\ \bibinfo {author} {\bibfnamefont
  {G.}~\bibnamefont {Torrieri}},\ }\href
  {https://doi.org/10.1103/PhysRevLett.111.172303} {\bibfield  {journal}
  {\bibinfo  {journal} {Phys. Rev. Lett.}\ }\textbf {\bibinfo {volume} {111}},\
  \bibinfo {pages} {172303} (\bibinfo {year} {2013})}\BibitemShut {NoStop}%
\bibitem [{\citenamefont {Strickland}(2015)}]{STRICKLAND2015}%
  \BibitemOpen
  \bibfield  {author} {\bibinfo {author} {\bibfnamefont {M.}~\bibnamefont
  {Strickland}},\ }\href {https://doi.org/10.1007/s12043-015-0972-1} {\bibfield
   {journal} {\bibinfo  {journal} {Pramana}\ }\textbf {\bibinfo {volume}
  {84}},\ \bibinfo {pages} {671} (\bibinfo {year} {2015})}\BibitemShut
  {NoStop}%
\bibitem [{\citenamefont {Nachman}\ and\ \citenamefont
  {Mangano}(2018)}]{Nachman2018}%
  \BibitemOpen
  \bibfield  {author} {\bibinfo {author} {\bibfnamefont {B.}~\bibnamefont
  {Nachman}}\ and\ \bibinfo {author} {\bibfnamefont {M.~L.}\ \bibnamefont
  {Mangano}},\ }\href {https://doi.org/10.1140/epjc/s10052-018-5826-9}
  {\bibfield  {journal} {\bibinfo  {journal} {The European Physical Journal C}\
  }\textbf {\bibinfo {volume} {78}},\ \bibinfo {pages} {343} (\bibinfo {year}
  {2018})}\BibitemShut {NoStop}%
\bibitem [{\citenamefont {Dirac}(1949)}]{RevModPhys.21.392}%
  \BibitemOpen
  \bibfield  {author} {\bibinfo {author} {\bibfnamefont {P.~A.~M.}\
  \bibnamefont {Dirac}},\ }\href {https://doi.org/10.1103/RevModPhys.21.392}
  {\bibfield  {journal} {\bibinfo  {journal} {Rev. Mod. Phys.}\ }\textbf
  {\bibinfo {volume} {21}},\ \bibinfo {pages} {392} (\bibinfo {year}
  {1949})}\BibitemShut {NoStop}%
\bibitem [{\citenamefont {Sjöstrand}\ \emph {et~al.}(2006)\citenamefont
  {Sjöstrand}, \citenamefont {Mrenna},\ and\ \citenamefont
  {Skands}}]{Sj_strand_2006}%
  \BibitemOpen
  \bibfield  {author} {\bibinfo {author} {\bibfnamefont {T.}~\bibnamefont
  {Sjöstrand}}, \bibinfo {author} {\bibfnamefont {S.}~\bibnamefont {Mrenna}},\
  and\ \bibinfo {author} {\bibfnamefont {P.}~\bibnamefont {Skands}},\ }\href
  {https://doi.org/10.1088/1126-6708/2006/05/026} {\bibfield  {journal}
  {\bibinfo  {journal} {Journal of High Energy Physics}\ }\textbf {\bibinfo
  {volume} {2006}},\ \bibinfo {pages} {026} (\bibinfo {year}
  {2006})}\BibitemShut {NoStop}%
\bibitem [{\citenamefont {Sjöstrand}\ \emph {et~al.}(2008)\citenamefont
  {Sjöstrand}, \citenamefont {Mrenna},\ and\ \citenamefont
  {Skands}}]{SJOSTRAND2008852}%
  \BibitemOpen
  \bibfield  {author} {\bibinfo {author} {\bibfnamefont {T.}~\bibnamefont
  {Sjöstrand}}, \bibinfo {author} {\bibfnamefont {S.}~\bibnamefont {Mrenna}},\
  and\ \bibinfo {author} {\bibfnamefont {P.}~\bibnamefont {Skands}},\ }\href
  {https://doi.org/https://doi.org/10.1016/j.cpc.2008.01.036} {\bibfield
  {journal} {\bibinfo  {journal} {Comput. Phys. Comm.}\ }\textbf {\bibinfo
  {volume} {178}},\ \bibinfo {pages} {852} (\bibinfo {year}
  {2008})}\BibitemShut {NoStop}%
\bibitem [{\citenamefont {Abesalashvili}\ \emph {et~al.}(1978)\citenamefont
  {Abesalashvili} \emph {et~al.}}]{Garsevanishvili78}%
  \BibitemOpen
  \bibfield  {author} {\bibinfo {author} {\bibfnamefont {L.~N.}\ \bibnamefont
  {Abesalashvili}} \emph {et~al.},\ }\href
  {http://www.jetpletters.ac.ru/ps/1573/article_24115.shtml} {\bibfield
  {journal} {\bibinfo  {journal} {JETP Lett.}\ }\textbf {\bibinfo {volume}
  {28}},\ \bibinfo {pages} {162} (\bibinfo {year} {1978})}\BibitemShut
  {NoStop}%
\bibitem [{\citenamefont {Abesalashvili}\ \emph {et~al.}(1979)\citenamefont
  {Abesalashvili} \emph {et~al.}}]{Garsevanishvili79}%
  \BibitemOpen
  \bibfield  {author} {\bibinfo {author} {\bibfnamefont {L.~N.}\ \bibnamefont
  {Abesalashvili}} \emph {et~al.},\ }\href
  {http://www.jetpletters.ac.ru/ps/1366/article_20666.shtml} {\bibfield
  {journal} {\bibinfo  {journal} {JETP Lett.}\ }\textbf {\bibinfo {volume}
  {30}},\ \bibinfo {pages} {448} (\bibinfo {year} {1979})}\BibitemShut
  {NoStop}%
\bibitem [{\citenamefont {Amaglobeli}\ \emph {et~al.}(1999)\citenamefont
  {Amaglobeli} \emph {et~al.}}]{Amaglobeli99}%
  \BibitemOpen
  \bibfield  {author} {\bibinfo {author} {\bibfnamefont {N.~S.}\ \bibnamefont
  {Amaglobeli}} \emph {et~al.},\ }\href
  {https://link.springer.com/article/10.1007/s100529901056} {\bibfield
  {journal} {\bibinfo  {journal} {Eur. Phys. J. C}\ }\textbf {\bibinfo {volume}
  {8}},\ \bibinfo {pages} {603} (\bibinfo {year} {1999})}\BibitemShut {NoStop}%
\bibitem [{\citenamefont {Djobava}\ \emph {et~al.}(2003)\citenamefont {Djobava}
  \emph {et~al.}}]{Djobava03}%
  \BibitemOpen
  \bibfield  {author} {\bibinfo {author} {\bibfnamefont {T.}~\bibnamefont
  {Djobava}} \emph {et~al.},\ }\href
  {http://www1.jinr.ru/Archive/Pepan/v-34-4/v-34-4-5.pdf} {\bibfield  {journal}
  {\bibinfo  {journal} {Phys. Part. Nucl.}\ }\textbf {\bibinfo {volume} {34}},\
  \bibinfo {pages} {526} (\bibinfo {year} {2003})}\BibitemShut {NoStop}%
\bibitem [{\citenamefont {Chkhaidze}\ \emph {et~al.}(2006)\citenamefont
  {Chkhaidze} \emph {et~al.}}]{Chkhaidze2006}%
  \BibitemOpen
  \bibfield  {author} {\bibinfo {author} {\bibfnamefont {L.}~\bibnamefont
  {Chkhaidze}} \emph {et~al.},\ }\href
  {https://inis.iaea.org/collection/NCLCollectionStore/_Public/38/063/38063548.pdf?r=1}
  {\bibinfo {title} {Light front variables in high energy inclusive reactions}}
  (\bibinfo {year} {2006}),\ \bibinfo {note} {proceedings of the Fifth Nuclear
  and Particle Physics Conference (NUPPAC-2005)}\BibitemShut {NoStop}%
\bibitem [{\citenamefont {Bleicher}\ \emph {et~al.}(1999)\citenamefont
  {Bleicher} \emph {et~al.}}]{UrQMD1}%
  \BibitemOpen
  \bibfield  {author} {\bibinfo {author} {\bibfnamefont {M.}~\bibnamefont
  {Bleicher}} \emph {et~al.},\ }\href
  {https://doi.org/10.1088/0954-3899/25/9/308} {\bibfield  {journal} {\bibinfo
  {journal} {J. Phys. G: Nucl. Phys}\ }\textbf {\bibinfo {volume} {25}},\
  \bibinfo {pages} {1859} (\bibinfo {year} {1999})}\BibitemShut {NoStop}%
\bibitem [{\citenamefont {Bass}\ \emph {et~al.}(1998)\citenamefont {Bass} \emph
  {et~al.}}]{UrQMD2}%
  \BibitemOpen
  \bibfield  {author} {\bibinfo {author} {\bibfnamefont {S.}~\bibnamefont
  {Bass}} \emph {et~al.},\ }\href
  {https://doi.org/https://doi.org/10.1016/S0146-6410(98)00058-1} {\bibfield
  {journal} {\bibinfo  {journal} {Prog. Part. Nucl. Phys.}\ }\textbf {\bibinfo
  {volume} {41}},\ \bibinfo {pages} {255 } (\bibinfo {year}
  {1998})}\BibitemShut {NoStop}%
\bibitem [{\citenamefont {Nair}(2021{\natexlab{a}})}]{nair2021light}%
  \BibitemOpen
  \bibfield  {author} {\bibinfo {author} {\bibfnamefont {R.~R.}\ \bibnamefont
  {Nair}},\ }\href@noop {} {\bibinfo {title} {On the light front surfaces in
  the phase space of hadrons in heavy-ion collisions}} (\bibinfo {year}
  {2021}{\natexlab{a}}),\ \Eprint {https://arxiv.org/abs/2103.04299}
  {arXiv:2103.04299 [hep-ph]} \BibitemShut {NoStop}%
\bibitem [{\citenamefont {Nair}(2021{\natexlab{b}})}]{nair2021feasibility}%
  \BibitemOpen
  \bibfield  {author} {\bibinfo {author} {\bibfnamefont {R.~R.}\ \bibnamefont
  {Nair}},\ }\href@noop {} {\bibinfo {title} {A feasibility study of the light
  front analysis of ultrarelativistic nuclear collisions in {ALICE} at {LHC}}}
  (\bibinfo {year} {2021}{\natexlab{b}}),\ \Eprint
  {https://arxiv.org/abs/2103.04932} {arXiv:2103.04932 [hep-ph]} \BibitemShut
  {NoStop}%
\bibitem [{\citenamefont {Nair}(2021{\natexlab{c}})}]{nair2021polynomial}%
  \BibitemOpen
  \bibfield  {author} {\bibinfo {author} {\bibfnamefont {R.~R.}\ \bibnamefont
  {Nair}},\ }\href@noop {} {\bibinfo {title} {A polynomial in transverse
  momentum manifesting thermalisation in nuclear collisions}} (\bibinfo {year}
  {2021}{\natexlab{c}}),\ \Eprint {https://arxiv.org/abs/2103.08190}
  {arXiv:2103.08190 [nucl-th]} \BibitemShut {NoStop}%
\bibitem [{\citenamefont {Andersson}\ \emph {et~al.}(1983)\citenamefont
  {Andersson}, \citenamefont {Gustafson}, \citenamefont {Ingelman},\ and\
  \citenamefont {Sjöstrand}}]{ANDERSSON198331}%
  \BibitemOpen
  \bibfield  {author} {\bibinfo {author} {\bibfnamefont {B.}~\bibnamefont
  {Andersson}}, \bibinfo {author} {\bibfnamefont {G.}~\bibnamefont
  {Gustafson}}, \bibinfo {author} {\bibfnamefont {G.}~\bibnamefont
  {Ingelman}},\ and\ \bibinfo {author} {\bibfnamefont {T.}~\bibnamefont
  {Sjöstrand}},\ }\href
  {https://doi.org/https://doi.org/10.1016/0370-1573(83)90080-7} {\bibfield
  {journal} {\bibinfo  {journal} {Physics Reports}\ }\textbf {\bibinfo {volume}
  {97}},\ \bibinfo {pages} {31} (\bibinfo {year} {1983})}\BibitemShut {NoStop}%
\bibitem [{\citenamefont {Sjöstrand}(1984)}]{SJOSTRAND1984469}%
  \BibitemOpen
  \bibfield  {author} {\bibinfo {author} {\bibfnamefont {T.}~\bibnamefont
  {Sjöstrand}},\ }\href
  {https://doi.org/https://doi.org/10.1016/0550-3213(84)90607-2} {\bibfield
  {journal} {\bibinfo  {journal} {Nuclear Physics B}\ }\textbf {\bibinfo
  {volume} {248}},\ \bibinfo {pages} {469} (\bibinfo {year}
  {1984})}\BibitemShut {NoStop}%
\bibitem [{\citenamefont {Sjöstrand}(2013)}]{sjostrand2013colour}%
  \BibitemOpen
  \bibfield  {author} {\bibinfo {author} {\bibfnamefont {T.}~\bibnamefont
  {Sjöstrand}},\ }\href@noop {} {\bibinfo {title} {Colour reconnection and its
  effects on precise measurements at the lhc}} (\bibinfo {year} {2013}),\
  \Eprint {https://arxiv.org/abs/1310.8073} {arXiv:1310.8073 [hep-ph]}
  \BibitemShut {NoStop}%
\bibitem [{\citenamefont {Ortiz~Velasquez}\ \emph {et~al.}(2013)\citenamefont
  {Ortiz~Velasquez}, \citenamefont {Christiansen}, \citenamefont
  {Cuautle~Flores}, \citenamefont {Maldonado~Cervantes},\ and\ \citenamefont
  {Pai\ifmmode~\acute{c}\else \'{c}\fi{}}}]{PhysRevLett.111.042001}%
  \BibitemOpen
  \bibfield  {author} {\bibinfo {author} {\bibfnamefont {A.}~\bibnamefont
  {Ortiz~Velasquez}}, \bibinfo {author} {\bibfnamefont {P.}~\bibnamefont
  {Christiansen}}, \bibinfo {author} {\bibfnamefont {E.}~\bibnamefont
  {Cuautle~Flores}}, \bibinfo {author} {\bibfnamefont {I.~A.}\ \bibnamefont
  {Maldonado~Cervantes}},\ and\ \bibinfo {author} {\bibfnamefont
  {G.}~\bibnamefont {Pai\ifmmode~\acute{c}\else \'{c}\fi{}}},\ }\href
  {https://doi.org/10.1103/PhysRevLett.111.042001} {\bibfield  {journal}
  {\bibinfo  {journal} {Phys. Rev. Lett.}\ }\textbf {\bibinfo {volume} {111}},\
  \bibinfo {pages} {042001} (\bibinfo {year} {2013})}\BibitemShut {NoStop}%
\bibitem [{\citenamefont {Ortiz}\ \emph {et~al.}(2015)\citenamefont {Ortiz},
  \citenamefont {Cuautle},\ and\ \citenamefont {Paić}}]{ORTIZ201578}%
  \BibitemOpen
  \bibfield  {author} {\bibinfo {author} {\bibfnamefont {A.}~\bibnamefont
  {Ortiz}}, \bibinfo {author} {\bibfnamefont {E.}~\bibnamefont {Cuautle}},\
  and\ \bibinfo {author} {\bibfnamefont {G.}~\bibnamefont {Paić}},\ }\href
  {https://doi.org/https://doi.org/10.1016/j.nuclphysa.2015.05.010} {\bibfield
  {journal} {\bibinfo  {journal} {Nuclear Physics A}\ }\textbf {\bibinfo
  {volume} {941}},\ \bibinfo {pages} {78} (\bibinfo {year} {2015})}\BibitemShut
  {NoStop}%
\bibitem [{\citenamefont {Abelev}\ \emph {et~al.}(2012)\citenamefont {Abelev}
  \emph {et~al.}}]{2012165}%
  \BibitemOpen
  \bibfield  {author} {\bibinfo {author} {\bibfnamefont {B.}~\bibnamefont
  {Abelev}} \emph {et~al.} (\bibinfo {collaboration} {ALICE collaboration}),\
  }\href {https://doi.org/https://doi.org/10.1016/j.physletb.2012.04.052}
  {\bibfield  {journal} {\bibinfo  {journal} {Physics Letters B}\ }\textbf
  {\bibinfo {volume} {712}},\ \bibinfo {pages} {165} (\bibinfo {year}
  {2012})}\BibitemShut {NoStop}%
\bibitem [{\citenamefont {Thakur}\ \emph {et~al.}(2018)\citenamefont {Thakur},
  \citenamefont {De}, \citenamefont {Sahoo},\ and\ \citenamefont
  {Dansana}}]{PhysRevD.97.094002}%
  \BibitemOpen
  \bibfield  {author} {\bibinfo {author} {\bibfnamefont {D.}~\bibnamefont
  {Thakur}}, \bibinfo {author} {\bibfnamefont {S.}~\bibnamefont {De}}, \bibinfo
  {author} {\bibfnamefont {R.}~\bibnamefont {Sahoo}},\ and\ \bibinfo {author}
  {\bibfnamefont {S.}~\bibnamefont {Dansana}},\ }\href
  {https://doi.org/10.1103/PhysRevD.97.094002} {\bibfield  {journal} {\bibinfo
  {journal} {Phys. Rev. D}\ }\textbf {\bibinfo {volume} {97}},\ \bibinfo
  {pages} {094002} (\bibinfo {year} {2018})}\BibitemShut {NoStop}%
\bibitem [{\citenamefont {Abazov}\ \emph {et~al.}(2010)\citenamefont {Abazov}
  \emph {et~al.}}]{PhysRevD.81.052012}%
  \BibitemOpen
  \bibfield  {author} {\bibinfo {author} {\bibfnamefont {V.~M.}\ \bibnamefont
  {Abazov}} \emph {et~al.} (\bibinfo {collaboration} {The D0 Collaboration}),\
  }\href {https://doi.org/10.1103/PhysRevD.81.052012} {\bibfield  {journal}
  {\bibinfo  {journal} {Phys. Rev. D}\ }\textbf {\bibinfo {volume} {81}},\
  \bibinfo {pages} {052012} (\bibinfo {year} {2010})}\BibitemShut {NoStop}%
\bibitem [{\citenamefont {Chekanov}\ \emph {et~al.}(2008)\citenamefont
  {Chekanov} \emph {et~al.}}]{20081}%
  \BibitemOpen
  \bibfield  {author} {\bibinfo {author} {\bibfnamefont {S.}~\bibnamefont
  {Chekanov}} \emph {et~al.} (\bibinfo {collaboration} {ZEUS Collaboration}),\
  }\href {https://doi.org/https://doi.org/10.1016/j.nuclphysb.2007.08.021}
  {\bibfield  {journal} {\bibinfo  {journal} {Nuclear Physics B}\ }\textbf
  {\bibinfo {volume} {792}},\ \bibinfo {pages} {1} (\bibinfo {year}
  {2008})}\BibitemShut {NoStop}%
\bibitem [{\citenamefont {Bierlich}\ and\ \citenamefont
  {Christiansen}(2015)}]{PhysRevD.92.094010}%
  \BibitemOpen
  \bibfield  {author} {\bibinfo {author} {\bibfnamefont {C.}~\bibnamefont
  {Bierlich}}\ and\ \bibinfo {author} {\bibfnamefont {J.~R.}\ \bibnamefont
  {Christiansen}},\ }\href {https://doi.org/10.1103/PhysRevD.92.094010}
  {\bibfield  {journal} {\bibinfo  {journal} {Phys. Rev. D}\ }\textbf {\bibinfo
  {volume} {92}},\ \bibinfo {pages} {094010} (\bibinfo {year}
  {2015})}\BibitemShut {NoStop}%
\bibitem [{\citenamefont {Acconcia}\ \emph {et~al.}(2018)\citenamefont
  {Acconcia}, \citenamefont {Chinellato}, \citenamefont {de~Souza},
  \citenamefont {Takahashi}, \citenamefont {Torrieri},\ and\ \citenamefont
  {Markert}}]{PhysRevD.97.036010}%
  \BibitemOpen
  \bibfield  {author} {\bibinfo {author} {\bibfnamefont {R.}~\bibnamefont
  {Acconcia}}, \bibinfo {author} {\bibfnamefont {D.~D.}\ \bibnamefont
  {Chinellato}}, \bibinfo {author} {\bibfnamefont {R.~D.}\ \bibnamefont
  {de~Souza}}, \bibinfo {author} {\bibfnamefont {J.}~\bibnamefont {Takahashi}},
  \bibinfo {author} {\bibfnamefont {G.}~\bibnamefont {Torrieri}},\ and\
  \bibinfo {author} {\bibfnamefont {C.}~\bibnamefont {Markert}},\ }\href
  {https://doi.org/10.1103/PhysRevD.97.036010} {\bibfield  {journal} {\bibinfo
  {journal} {Phys. Rev. D}\ }\textbf {\bibinfo {volume} {97}},\ \bibinfo
  {pages} {036010} (\bibinfo {year} {2018})}\BibitemShut {NoStop}%
\bibitem [{\citenamefont {Skands}\ \emph {et~al.}(2014)\citenamefont {Skands},
  \citenamefont {Carrazza},\ and\ \citenamefont {Rojo}}]{Skands_2014}%
  \BibitemOpen
  \bibfield  {author} {\bibinfo {author} {\bibfnamefont {P.}~\bibnamefont
  {Skands}}, \bibinfo {author} {\bibfnamefont {S.}~\bibnamefont {Carrazza}},\
  and\ \bibinfo {author} {\bibfnamefont {J.}~\bibnamefont {Rojo}},\ }\bibfield
  {journal} {\bibinfo  {journal} {The European Physical Journal C}\ }\textbf
  {\bibinfo {volume} {74}},\ \href
  {https://doi.org/10.1140/epjc/s10052-014-3024-y}
  {10.1140/epjc/s10052-014-3024-y} (\bibinfo {year} {2014})\BibitemShut
  {NoStop}%
\bibitem [{\citenamefont {{The ALICE collaboration}}(2013)}]{V02013}%
  \BibitemOpen
  \bibfield  {author} {\bibinfo {author} {\bibnamefont {{The ALICE
  collaboration}}},\ }\href {https://doi.org/10.1088/1748-0221/8/10/p10016}
  {\bibfield  {journal} {\bibinfo  {journal} {JINST}\ }\textbf {\bibinfo
  {volume} {8}}\bibinfo  {number} { (10)},\ \bibinfo {pages}
  {P10016}}\BibitemShut {NoStop}%
\bibitem [{\citenamefont {{The ALICE
  collaboration}}(2014)}]{doi:10.1142/S0217751X14300440}%
  \BibitemOpen
\bibfield  {number} {  }\bibfield  {author} {\bibinfo {author} {\bibnamefont
  {{The ALICE collaboration}}},\ }\href
  {https://doi.org/10.1142/S0217751X14300440} {\bibfield  {journal} {\bibinfo
  {journal} {International Journal of Modern Physics A}\ }\textbf {\bibinfo
  {volume} {29}},\ \bibinfo {pages} {1430044} (\bibinfo {year}
  {2014})}\BibitemShut {NoStop}%
\bibitem [{\citenamefont {Khuntia}\ \emph {et~al.}(2021)\citenamefont {Khuntia}
  \emph {et~al.}}]{Khuntia_2021}%
  \BibitemOpen
  \bibfield  {author} {\bibinfo {author} {\bibfnamefont {A.}~\bibnamefont
  {Khuntia}} \emph {et~al.},\ }\href {https://doi.org/10.1088/1361-6471/abb1f8}
  {\bibfield  {journal} {\bibinfo  {journal} {J. Phys. G: Nucl. Phys}\ }\textbf
  {\bibinfo {volume} {48}},\ \bibinfo {pages} {035102} (\bibinfo {year}
  {2021})}\BibitemShut {NoStop}%
\bibitem [{\citenamefont {Brun}\ and\ \citenamefont
  {Rademakers}(1997)}]{BRUN199781}%
  \BibitemOpen
  \bibfield  {author} {\bibinfo {author} {\bibfnamefont {R.}~\bibnamefont
  {Brun}}\ and\ \bibinfo {author} {\bibfnamefont {F.}~\bibnamefont
  {Rademakers}},\ }\href {https://doi.org/10.1016/S0168-9002(97)00048-X}
  {\bibfield  {journal} {\bibinfo  {journal} {Nucl. Inst. and Meth. in Phys.
  Res. A}\ }\textbf {\bibinfo {volume} {389}},\ \bibinfo {pages} {81} (\bibinfo
  {year} {1997})}\BibitemShut {NoStop}%
\end{thebibliography}%
\end{document}